\documentclass[aps,prl,floats,twocolumn,showpacs]{revtex4}
\usepackage{epsfig}
\usepackage{graphicx}
\usepackage{rotate}
\begin{document}
\topmargin 0.5cm.
\title{Amplified signal response in scale-free networks by collaborative signaling}

\author{Juan A. Acebr\'on}

\affiliation{Departament d'Enginyeria Inform{\`a}tica i Matem{\`a}tiques,
  Universitat Rovira i Virgili, 43007 Tarragona, Spain}

\author{Sergi Lozano}

\affiliation{Departament d'Enginyeria Inform{\`a}tica i Matem{\`a}tiques,
  Universitat Rovira i Virgili, 43007 Tarragona, Spain}

\author{Alex Arenas}

\affiliation{Departament d'Enginyeria Inform{\`a}tica i Matem{\`a}tiques,
  Universitat Rovira i Virgili, 43007 Tarragona, Spain}

\date{\today}

\begin{abstract}

Many natural and artificial two-states signaling devices are connected forming networks.
The information-processing potential of these systems is usually related to the response to weak external signals. Here, using a network of overdamped bistable elements, we study the effect of a heterogeneous complex topology on the signal response. The analysis of the problem in random scale-free networks, reveals that heterogeneity plays a crucial role in amplifying external signals. We have contrasted numerical simulations with analytical calculations in simplified topologies.

 \end{abstract}

\maketitle

Signaling devices are the building blocks of many natural and artificial information-processing systems, for example, cells in living organisms respond to their environment by means of an interconnected network of receptors, messengers, protein kinases and other signaling molecules \cite{bray,alon}. In artificial systems, we remark recent studies of networks of fluxgate magnetometers, with potential ways to enhance the utility and sensitivity of a large class of nonlinear dynamic sensors e.g. the magnetometers, ferroelectric detectors for electric fields, or piezoelectric detectors for acoustics applications by careful coupling and configuration \cite{abbs,vb,bl}. Their proper functioning always imply high sensitivity to external signals.

One of the most remarkable discoveries in non-linear physics, over the last thirty years, was the phenomenon of stochastic resonance. It represents the surprising effect manifested in nonlinear dynamics where a weak subthreshold input signal can be amplified by the assistance of noise. The requirements for such amplification are three: (i) a non-linear dynamical system endowing a potential with energetic activation barriers, (ii) a small amplitude (usually periodic) external signal, and (iii) a source of noise inherent or coupled to the system. Given these features, the response of the system undergoes resonance-like behavior as a function of the noise floor; hence the name stochastic resonance \cite{benzi, Hanggi}.

The response of the system can be enhanced by coupling it through
all-to-all \cite{morillo} or in arrays configurations \cite
{Bulsara1}. This has been done mainly resorting to a linear coupling
among oscillators. The most intriguing part of the phenomenon is
that noise (a source of disorder) can favor the amplification of a
signal, an effect that usually requires a very precise synchrony
(order). Recently in \cite{Tessone}, it has been shown that even in
absence of noise a similar beneficial effects can be observed.
There, different sources of diversity plays the role of the noise,
inducing a resonant collective behavior. Moreover, it was pointed
out that such a resonance effect does not depend on the source of
disorder, and could be observed even in nonregular network of
connectivities.

Our goal here is to give evidences that an equivalent amplification of a external signal can be obtained in a deterministic dynamical system of signaling devices in complex heterogeneous networks.

The typical example, where the phenomenon has been investigated, consists on a bistable potential with a periodic signal in a thermal bath. A general dimensionless equation describing this scenario is
\begin{equation}
\dot x = x -x^3 +A\sin(\omega t) +\eta(t)
\label{SR}
\end{equation}
\noindent where the bistable potential is $V(x)= -x^{2}/2+x^{4}/4$, the external signal $A\sin(\omega t)$ and $\eta(t)$ a Gaussian white noise with zero mean and autocorrelation $\langle \eta(t)\eta(t')\rangle=2D\delta(t-t')$. Eq.(\ref{SR}) represents the overdamped motion of a Brownian particle in a bistable potential and periodic forcing. In absence of forcing and noise, the system has two stable fixed points centered around $\pm 1$, which corresponds to the minimum of the potential energy function $V(x)$. When the amplitude of the forced signal is subthreshold, each node $i$ oscillates around the minimum of its potential with the same frequency of the forcing signal.

The deterministic system we will study corresponds to a network of elements obeying Eq.(\ref{SR}) in absence of noise. Noise has been explicitly excluded to avoid a possible uncontrolled superposition of effects, that would make difficult to determine the influence of the topology. The network is expressed by its adjacency matrix $A_{ij}$ with entries 1 if $i$ is connected to $j$, and 0 otherwise. For simplicity, from now on we consider only undirected unweighted networks. Mathematically the system reads
\begin{equation}
\dot x_i = x_i -x_{i}^3 +A\sin(\omega t) +\lambda L_{ij}x_j \hspace{0.5cm}
i=1,...,N,\label{SRL}
\end{equation}
where $L_{ij}=k_i\delta_{ij} - A_{ij}$ is the Laplacian matrix of the network, being $k_{i}=\sum_j A_{ij}$ the degree of node $i$.

\begin{figure}
\epsfig{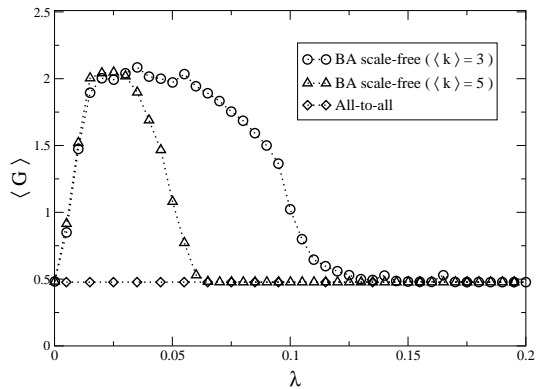} \caption{Average amplification $\langle G\rangle$ as a function of the coupling $\lambda$, for both topologies, scale-free and all-to-all. Results corresponding to two different values of the average degree ($\langle k\rangle=3$, marked with circles, and $\langle k\rangle=5$ with triangles) for the BA model network are plotted.  Here $N=500$, and $\omega=2\pi 10^{-1}$. The simulations were averaged over $1000$ randomly chosen initial conditions. While the scale-free networks show a clear amplification of the external signal, for a significative range of values of $\lambda$ that depends on the average degree, the all-to-all connectivity does not amplifiy the signal.}
\label{ER-SF}
\end{figure}

We have conducted numerical simulations of the system above for the Barabasi-Albert (BA) network model, and all-to-all connectivity networks. While BA has become the paradigm of a growing model that provides with a scale-free (power-law) degree distribution \cite{barabasi}, all-to-all connectivity is a representative of homogeneous in degree networks.\\
The results are depicted in Figure \ref{ER-SF}. Here
the average amplification $\langle G\rangle$ over different initial conditions has been computed as function of the coupling
$\lambda$, being the amplification defined as $G=\max_i {x_i}/A$. The frequency of the forcing signal $\omega$, and the number of nodes $N$ were here kept fixed. The obtained results show that in BA networks an amplification of the signal occurs, while in all-to-all networks no amplification of the external signal is observed for any value of the coupling $\lambda$.

This difference of behavior is attributed to the presence of hubs in the BA networks, which are the main responsible for the heterogeneity in degree of the network. Keeping this in mind, we study a topology consisting on a star-like network (one hub and N-1 peripheral leaves). Such a topology is simple enough to be mathematically tractable, and simultaneously
capable to capture the main trait of heterogeneity found in scale-free networks. Indeed, the resulting dynamical system
decomposes in two parts: the dynamics of the hub (the highly connected
node in the network with $N-1$ links), $x_H$,  and the leaves, $y_i$,
linked to the hub. Then, the dynamical
system (\ref{SRL}) becomes
\begin{equation}
\dot x_H = [1-\lambda (N-1)]x_{H} -x_{H}^3 +A\sin(\omega t)
+\lambda\sum_{i=1}^{N-1} y_i \label{ESFH}
\end{equation}
\begin{equation}
 \dot y_i = (1-\lambda)y_{i} -y_{i}^3 +A\sin(\omega t) + \lambda x_H, \hspace{0.5cm}
i=1,...,N \label{ESFHH}
\end{equation}
Now we make the following hypothesis: for a coupling $\lambda$
sufficiently small, the dynamics of the leaves can be decoupled from
that of the hub, obtaining then from Eq.(\ref{ESFHH}) the well known
equation of an overdamped bistable oscillator. {\bf Moreover, when
the signal is weak enough compared with the potential barrier, this
can be conveniently linearized around one of the potential minima}.
Then, Eq.(\ref{ESFHH}) can be solved for the $i$th node, and
asymptotically for long time yields,
\begin{equation}
y_i(t)_{ t\rightarrow\infty}\sim \xi_i-\frac{A}{\omega^2+4}\left[\omega \cos
(\omega t)-2 \sin (\omega t)\right]
\label{y_dyn}
\end{equation}
with  $\xi_i=\pm 1$ depending on the initial conditions.
Inserting this solution into Eq.(\ref{ESFH}) we obtain
\begin{equation}
\dot x_H = -V'_H (x_H) +A'\sin(\omega t) +B' \cos (\omega t)+\lambda \eta_i
\label{hub_dyn}
\end{equation}
where
\begin{equation}
V_H(x)=-[1-\lambda (N-1)]x^2/2 +x^4/4
\label{pot_hub}
\end{equation}
is the effective potential felt by the hub. The rest of parameters in Eq.(\ref{hub_dyn}) are
$\eta_i= \sum_{i=1}^{N-1} \xi_i$, and
$A'=A[1+2\lambda (N-1)]/(\omega^2+4)$, $B'=-2\lambda (N-1)/(\omega^2+4)$.
Notice that, the problem has been reduced to the motion of an overdamped oscillator, in an effective potential driven by a reamplified forcing signal coming from the global sum of the leaves. The two possible solutions for the nodes Eq.(\ref{y_dyn}), namely oscillations around $\pm 1$, affect the dynamics of the hub as a quenched disorder represented by $\lambda \eta_i$, and is equivalent to the diversity studied in \cite{Tessone}. Strictly speaking, the central limit theorem allows us to state that in the limit $N\to\infty$, $\eta_i$ behaves as a random variable governed by a gaussian probability distribution with variance $\sigma^2=N-1$. This is so because the initial conditions were randomly chosen.
The height  of the potential barrier for the hub is now $h=[1-(N-1)\lambda]^2/4$, and induces a strongly dependence on $\lambda$ via the factor $(N-1)$. Moreover, when $\lambda=1/(N-1)$ the barrier for the hub disappears, leading to a unique single fixed point $X_H=0$. Note that both mechanisms, the quenched disorder and the decrement of the potential barrier, rooted on the heterogeneity of the network, may cooperate to allow the hub to surmount the potential barrier. When $\lambda\neq 1/(N-1)$, {\bf  and for coupling and signal amplitude sufficiently small}, Eq.(\ref{hub_dyn}) can be solved in the asymptotic
long-time limit and yields
\begin{widetext}
\begin{equation}
\centering{ x_H(t, \eta )_{t\rightarrow\infty} \sim x_H^{(0)}
-\frac{1}{\omega^2+a_H^2}\left[(B' \omega-A'\, a_H)
\sin (\omega t)- (B'\,a_H+A'\omega)
\cos (\omega t)
 \right] }
\label{solH}
\end{equation}
\end{widetext}
where $a_H=V''_H (x_H^{(0)})$ being $x_H^{(0)}$ the equilibrium points in absence of forcing. Note that $x_H(t, \eta)$ depends also implicitly on the random variable $\eta$ through the equilibrium point $x_H^{(0)}$ to which the hub dynamics relaxes.
Knowing the long time evolution for the hub, its amplification can be readily evaluated, and the result is
\begin{equation}
G(\eta)= \frac{1}{A}\frac{1}{a_H^2+\omega^2}
\sqrt{(B' \omega -A'\, a_H)^2
+(B' \,a_H +A'\, \omega)^2
}
\label{ESFN}
\end{equation}
\begin{figure}
\epsfig{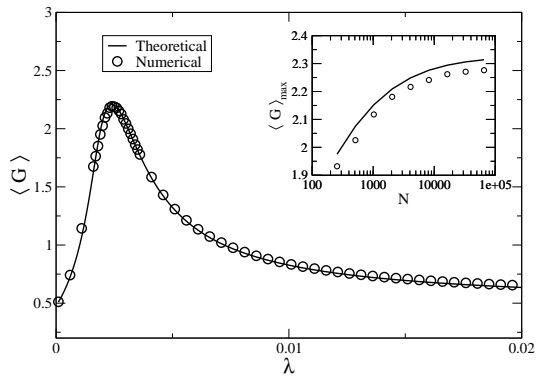}
\caption{Average amplification $\langle G\rangle$ as a function of the coupling $\lambda$. We compare the numerical simulation results of the dynamics of the star-like network Eqs.(\ref{ESFH}-\ref{ESFHH}), and the theoretical result Eq.(\ref{amplificationteo}). Inset:  Maximum average amplification $\langle G\rangle_{max}$ as a function of the size of the network $N$. We compare again, the numerical simulation results of the dynamics of the star-like network Eqs.(\ref{ESFH}-\ref{ESFHH}), and the theoretical approximation Eq.(\ref{asympto}). Parameters are as in Fig.~1}
\label{fig_model}
\end{figure}

In practice, we must average over initial conditions to cancel out the dependence
on them. However, this turns out to be equivalent to averaging over $\eta_i$. Therefore, the average amplification is given by
\begin{equation}
\langle G\rangle=\frac{1}{\sqrt{2\pi (N-1)}}\int_{-\infty}^{\infty}d\eta\, e^{-\frac{\eta^2}{2(N-1)}}
G(\eta)
\label{amplificationteo}
\end{equation}
The influence on the size of the network, $N$ can be analyzed resorting to the equation above. In particular, we are interested in the case of $N$ large, and for $\lambda\approx1/(N-1)$, when the average amplification attains its maximum value. Making use of the Laplace method, it is possible to find an asymptotic solution to Eq.(\ref{amplificationteo}), in the limit $N\to\infty$, which is
\begin{equation}
\langle G\rangle=G(0)+\frac{1}{N}\left.\frac{d^2 G}{d\eta^2}\right|_{\eta=0}\eta+O(N^{-2}),
\label{asympto}
\end{equation}
Therefore, the maximum average amplification can be extracted from a network, turns out to be
 bounded from above by $G(0)$.

\begin{figure}
\epsfig{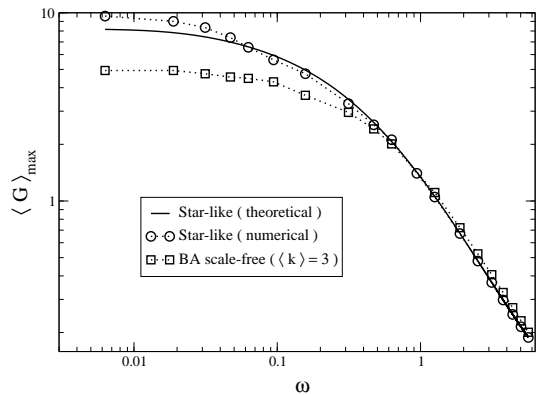} \caption{Maximum average amplification $\langle G\rangle_{max}$
obtained for the star-like network and BA network as a function of the frequency
of the external signal. The coupling is kept fixed to the value $\lambda =0.04$ for the BA network, and $\lambda=0.002$ for the star-like network. These coupling values are those that exhibit the maximum of  $\langle G\rangle$ in Fig.~1 and Fig.~2, respectively. the solid line corresponds to the theoretical prediction for the star-like network given by Eq.(\ref{amplificationteo}).
Other parameters are as in Fig.~1.} \label{fig_freq}
\end{figure}

We have performed extensive numerical
simulations for the dynamical system
(\ref{ESFH})-(\ref{ESFHH}), varying $N$ and $\lambda$. In Fig.~\ref{fig_model}, the results of the amplification in Eq.(\ref{amplificationteo}) and the numerical simulations
of the system equations are compared for different values of
the coupling $\lambda$, keeping fixed $N$ and $\omega$. The amplification grows monotonically with $\lambda$ until reaching a maximum value. Such a value coincides approximately with $\lambda=1/(N-1)$, which corresponds to the minimum value of
the barrier height. Therefore, the maximum signal amplification occurs when
the barrier of the effective potential of the hub attains its minimal value. This allows the hub to move back and forth between the two minima of the effective potential, yielding a much wider oscillation. It is important to observe that
a similar beneficial effect has been found in the stochastic resonance
phenomena, but rather here it is due merely to the natural heterogeneity of the network. Then, it can be seen as a kind of resonance induced by topology, i.e. a sort of {\em topological resonance}.
The inset
shows the maximum amplification obtained for different values of $N$. Notice that for large $N$, the solution
tends asymptotically to a constant value as expected from Eq.(\ref{asympto}).
Finally, in Fig.~\ref{fig_freq}, we compare
the maximum amplification in different networks as a function of  the frequency of the external signal. The amplification decreases monotonically with the frequency as predicted by the model.

The remarkable agreement between simulations and analytical results validates our hypothesis of decoupling the hub from leaves. We are now in the situation of interpreting the results in the BA scale-free networks. A neat picture is given by considering that each highly connected node acts locally as it were the hub of a star-like network, with a degree $k$ picked up from the degree distribution. Therefore, for a given coupling $\lambda$, we can find several star-like networks in different stages, depending on the degree of its local hubs. Recall that when $\lambda=1/k$, the maximum signal amplification is attained for a hub with degree $k$. Furthermore, increasingly larger values of $\lambda$ would activate local hubs with smaller degrees. Since the network has several hubs, it should exist a wide range of values of $\lambda$ for which the maximum amplification is achieved following a cascade of amplification. In Fig.~\ref{ER-SF} such a behavior can be clearly observed. This contrasts with the results found for the star-like network, and  anticipates that the scale-free network consists on a much more robust topology where such a new phenomena occurs.

\begin{figure}
\epsfig{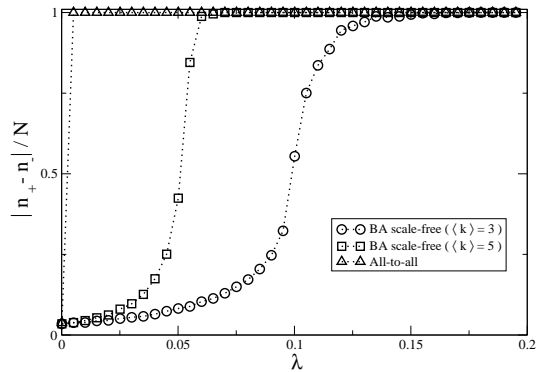} \caption{Degree
of synchronization of the BA and all-to-all networks versus the coupling. Two
different average degree values for the BA network are plotted. The larger the average degree, the lower the critical coupling needed to achieve full synchronization is. The all-to-all network synchronizes for any value of $\lambda\neq 0$.
Parameters are as in Fig.~1} \label{synchro}
\end{figure}

The mechanism described above remains valid until full
synchronization occurs. For large values of $\lambda$, the nodes
become strongly coupled, and then they spatially synchronize
behaving as a single node. We define the degree of spatial
synchronization as  the fraction of nodes $|n_{+}-n_{-}|/N$ where
$n_{+}$ and $n_{-}$ are the number of elements at the positive or
negative well, respectively. In Fig.~\ref{synchro} we show how full
synchronization emerges once a critical coupling is attained. In
addition, numerical simulations depicted in the same figure reveal
that the path to synchronization is more pronounced for scale-free
networks with larger average degree. Then, for this case a smaller
range of values for which maximum amplification is sustained should
be observed in accordance with Fig.~\ref{ER-SF}. Similar curves have
been found in the prototype Kuramoto model \cite{Acebron}, which
describes synchronization phenomena in nonlinear coupled
oscillators, on top of complex networks \cite{Arenas1}. However, the
relevance of the present case stems from the fact that
synchronization, contrary to which often happens in nonlinear
coupled dynamics, has a negative effect in the amplification
process. The all-to-all configuration in Fig.~\ref{synchro} presents
full synchronization for any $\lambda$ different from 0. The reason
for this observed behavior is that no source of disorder exist in
this case. The spatial symmetry of the problem is unstable, and any
small difference in concentration of elements in one of the two
wells produces an asymmetry in the potential that evolves towards a
monostable (synchronized) system. Note that this mechanism explains
also the absence of amplification in Fig.~\ref{ER-SF}.

Summarizing, in this work we have shown that a topological resonant-like effect emerges in scale-free topologies of signaling devices, exemplified by a deterministic overdamped bistable dynamical system.
The cooperative interaction of nodes connected to the hub due to the topology, has been shown to be the key feature for amplifying external signals. We have analytically proved that the phenomenon can be totally characterized in the framework of a simple topology consisting of a star-like network. The agreement between simulations and analytical predictions allow us to understand the amplification curves as a function of the parameters of the system: the coupling, the size of the system and the frequency of the periodic external signal. At the light of the current results, we speculate that the ubiquity of scale-free topological structures in biological systems can be related to its ability to amplify sensitivity to weak external signals.

\begin{acknowledgments}

This work is supported by
Spanish Ministry of Science and Technology Grant FIS2006-13321-C02. J.A.A. acknowledges support from the Ministerio de Ciencia y Tecnolog\'{\i}a (MEC)
through the Ram\'{o}n y Cajal programme. We acknowledge for the usage of the resources, technical expertise and assistance provided by BSC-CNS supercomputing facility.

\end{acknowledgments}

\end{document}